%Paper: hep-lat/9206010
%From: maarten@aapje.wustl.edu (Maarten Golterman)
%Date: Thu, 4 Jun 92 17:37:00 CDT

%&texsis
%%%%%%%%%%%%%%%%%%%%%%%%%%%%%%%%%%%%%%%%%%%%%%%%%%%%%%%%%%%%%%%%%%%%%%%%
%%%%%%%%%%%%%%%%%%%%%%%%%%%%%%%%%%%%%%%%%%%%%%%%%%%%%%%%%%%%%%%%%%%%%%%%
%%This file uses the Texsis macropackage, available from the database.%%
%%%%%%%%%%%%%%%%%%%%%%%%%%%%%%%%%%%%%%%%%%%%%%%%%%%%%%%%%%%%%%%%%%%%%%%%
%%%%%%%%%%%%%%%%%%%%%%%%%%%%%%%%%%%%%%%%%%%%%%%%%%%%%%%%%%%%%%%%%%%%%%%%
%{[(
%%%#[ Beginstuff:
\overfullrule=0pt  % uncomment this to turn off overfull boxes
\texsis
\preprint
%\draft
\Eurostyletrue
\superrefsfalse

\def\Acknowledgements{{\bigskip\leftline{{\bf Acknowledgments}}\medskip}}
%%%#] Beginstuff:
%%%#[ References:
\referencelist
\reference{EichPres}
E.~Eichten and J.~Preskill,
\journal Nucl.Phys. B;268,179(1986)
\endreference
\reference{KarstenSmit}
L.~H.~Karsten and J.~Smit,
\journal Nucl. Phys. B ;183,103(1981)
\endreference
\reference{Ginsparg}
see also P.~Ginsparg, thesis, 1981, unpublished
\endreference
\reference{SmitZak}
J.~Smit,
\journal Acta Physica Polonica;B17,531(1986)
\endreference
\reference{Swift}
P.~D.~V.~Swift,
\journal Phys. Lett. B;145,256(1984)
\endreference
\reference{Maiani}
A.~Borelli, L.~Maiani, G.-C.~Rossi, R.~Sisto and M.~Testa,
\journal Nucl. Phys. B;333,335(1990)
\endreference
\reference{Zaragoza}
J.L.~Alonso, Ph.~Boucaud, J.L.~Cortes and E.~Rivas,
Mod. Phys. Lett. A \underbar{5} (1990) 275;
Nucl. Phys. B {\rm (Proc. Suppl.)} \underbar{17} (1990) 461
\endreference
\reference{SmitSeillac}
J.~Smit,
\journal Nucl. Phys. B {\rm (Proc. Suppl.)};4,451(1988)
\endreference
\reference{BodwinKovacs}
G.~T.~Bodwin and E.~V.~Kovacs,
\journal Nucl. Phys. B {\rm (Proc. Suppl.)};20,546(1991)
\endreference
\reference{*BodwinKovacs}
preprint ANL-HEP-CP-91-75,
to be publ. in Proc. of Particles and Fields '91 Conference, Vancouver,
Canada, Aug 18-22, 1991
\endreference
\reference{AachenYukawa}
W.~Bock, A.~K.~De, C.~Frick, K.~Jansen and T.~Trappenberg,
\journal Nucl. Phys. B;371,683(1992)
\endreference
\reference{JanandUs}
M.~F.~L.~Golterman, D.~N.~Petcher and J.~Smit,
\journal Nucl. Phys. B ;370,51(1992)
\endreference
\reference{UsEPJapan}
M.~F.~L.~Golterman and D.~N.~Petcher,
preprint Wash.~U.~HEP/91-61,
to be published in the proceedings of the 1991 Conference
on Lattice Field Theory, KEK, Tsukuba, Japan
\endreference
\reference{Georgi}
H.~Georgi, Lie Algebras in Particle Physics, Benjamin Cummings
Publishing Company, 1982
\endreference
\reference{Banks}
T.~Banks,
\journal Phys. Lett. B;272,75(1991)
\endreference
\reference{Manohar}
M.~J.~Dugan and A.V.~Manohar,
\journal Phys. Lett. B;265,137(1991)
\endreference
\reference{Banksnew}
T.~Banks and A.~Dabholkar,
preprint RU-92-09(1992)
\endreference
%\reference{O(10)model}
%T.~Matsui, H.~Kleinert and S.~Ami,
%\journal Phys. Lett. B;143,199(1984)
%\endreference
\reference{Jantransfer}
J.~Smit,
\journal Nucl. Phys. B {\rm (Proc. Suppl.)};20,542(1991)
\endreference
\reference{Usoneoverw}
M.~F.~L.~Golterman and D.~N.~Petcher,
\journal Nucl. Phys. B {\rm (Proc. Suppl.)};20,577(1991)
\endreference
\reference{*Usoneoverw}
M.~F.~L.~Golterman, D.~N.~Petcher and E.~Rivas,
preprint Wash.~U.~HEP/91-82, to be published in Nucl. Phys.~B
\endreference
\reference{Usoneoverd}
M.~F.~L.~Golterman and D.~N.~Petcher,
\journal Phys. Lett. B;247,370(1990)
\endreference
\reference{*Usoneoverd}
\journal Nucl. Phys. B;359,91(1991)
\endreference
\reference{Aachenoneoverw}
W.~Bock, A.~K.~De and J.~Smit,
preprint HLRZ-91-81, J\"ulich and refs. therein
\endreference                   % Not listed as so in spires.
\reference{Kutietal}
A.~Hasenfratz, P.~Hasenfratz, K.~Jansen, J.~Kuti and Y.~Shen,
%preprint UCSD/PTH 91-06.
\journal Nucl. Phys. B;365,79(1991)
\endreference
\reference{Coleman}
{S.~Coleman,
in {\sl ``Pointlike Structures Inside and Outside Hadrons''}
(edited by A.~Zichichi)
{\sl Plenum Press, New York}
(1980)
{\it (Erice 1979)}}
\endreference
\reference{Aachenphase}
W.~Bock, A.~K.~De, K.~Jansen, J.~Jers\'ak, T.~Neuhaus and J.~Smit,
\journal Nucl. Phys. B;344,207(1990)
\endreference

\endreferencelist
%%%#] References:
%%%#[ Math macros:

\def\frac#1#2{{{\scriptstyle #1} \over {\scriptstyle #2}}}

\def\quart{{{\scriptscriptstyle\frac{1}{4}}}}
\def\unity{{1\!\!1}}

\def\ltilde{{\tilde\lambda}}
\def\rtilde{{\tilde{r}}}
\def\ytilde{{\tilde{y}}}
\def\wtilde{{\tilde{w}}}
\def\gtilde{{\tilde{g}}}
\def\vtilde{{\tilde{v}}}
\def\Itilde{{\tilde{I}}}
\def\Idagger{{I^\dagger}}
\def\I#1{{I_#1}}
\def\IY#1{{I_{Y#1}}}
\def\H#1#2{{H_#1^#2}}
\def\HY#1#2{{H_{Y#1}^#2}}

\def\psidagger{{\psi^\dagger}}
\def\Psidagger{{\Psi^\dagger}}
\def\psistar{{\psi^*}}
\def\psiT{\psi^T}
\def\dfermi{{ \lbrack d\psi\vphantom{\psidagger}\rbrack
             \lbrack d\psidagger\rbrack}}
\def\B{B}

\def\sigmahat{\hat\sigma}
\def\sigmabar{{\overline\sigma}}

\def\muhat{{\hat \mu}}
\def\Lag{{\cal L}}
\def\Lagkin{\Lag_{\psi^2}}

\def\SEP{{S_{\rm EP}}}
\def\LagEP{{\Lag_{\rm EP}}}
\def\Lagscalarkin{\Lag_\phi}
\def\Lagfour{\Lag_{\psi^4}}

\def\LaggenWilson{\Lag_{\rm GW}}
\def\LagY{\Lag_{Y}}
\def\LagEPY{\Lag^\prime_{\rm EP}}
\def\LagWY{\Lag_{\rm WY}}
\def\hc{\hbox{\rm h.c.}}

\def\D{{\partial}}

\def\Dmu{{\partial_\mu}}
\def\Dmutilde{{\tilde\D_\mu}}
\def\square#1#2{{\vcenter{\vbox{\hrule height.#2pt
    \hbox{\vrule width.#2pt height#1pt \kern#1pt \vrule width.#2pt}
    \hrule height.#2pt}}}}
\def\Box{{\mathchoice{\sl\square63}{\sl\square63}\square{2.1}3\square{1.5}3}\,}

%\def\Dg{{D}}
%\def\Dgtilde{{\tilde \Dg}}

%\def\epsL{{\epsilon_L}}

%\def\epsLbar{{\overline\epsilon_L}}
%\def\chibar{{\overline\chi}}

%\def\Fpropcinv{{S^{(c)-1}_F}}
%\def\Fpropcren{S^{(c)}_{F\hbox{\rm ren}}}
%\def\Fpropcreninv{S^{(c)-1}_{F\hbox{\rm ren}}}

%\def\Stilde{{\tilde S}}
%\def\Umunx{{V^{\dagger}(x)U_{\mu}(x) V(x+\hat{\mu})}}
%\def\phibar{{\overline\phi_{\rm (cont)}}}
%\def\Dgbar{{\overline\Dg^{\rm (cont)}}}
%\def\phibardagger{{\overline\phi^\dagger_{\rm (cont)}}}

%%%#] Math macros:
%%%#[ Title Page:
% preprint WASH-U-HEP 92/??  (in 60's (MFLG) or 80's (DP))
%\pubdate{April, 1992}
\pubcode{Wash. U. HEP/92-80}
\titlepage
\title
Absence of Chiral Fermions in the Eichten--Preskill Model
\endtitle
\author
Maarten~F.~L.~Golterman$^\dagger$, Donald~N.~Petcher$^\ddagger$ and
Elena~Rivas$^*$
Department of Physics
Washington University
St. Louis, MO 63130-4899
USA
\endauthor
\abstract

The model proposed by Eichten and Preskill for obtaining theories with
chiral fermions from the lattice is shown to undergo spontaneous
symmetry breaking. In addition, the fermions appear to be
Dirac-like everywhere in the phase diagram with no room for undoubled
Weyl fermions. The phase diagram of a closely related Higgs-Yukawa model
is similar to that of the
Smit--Swift model, which also does not give rise to chiral fermions.
These results cast serious doubts on the original
scenario for the emergence of chiral fermions.

\endabstract
\bigskip
\vfootnote{$\dagger$}{email address: maarten@wuphys.wustl.edu}
\vfootnote{$\ddagger$}{email address: petcher@wuphys.wustl.edu}
\vfootnote*{email address: elena@wuphys.wustl.edu}
\endtitlepage

%%%#] Title Page:
%%%#[ Introduction:

\section{Introduction}

One of the earliest proposals for obtaining a theory with chiral
fermions from the lattice was made in an elegant paper by Eichten and
Preskill\cite{EichPres} several years ago. The philosophy behind the
proposal is to start with a gauge theory containing only left-handed
chiral fermion fields, and to add gauge invariant `generalized Wilson'
terms to the lagrangian which vanish in the classical continuum limit
and are explicitly constructed to break all symmetries of the model that
are broken through anomalies in the target continuum theory. In
addition, the new terms must be carefully chosen so as to be irrelevant
with regard to the particles one desires to survive in the continuum
theory, but they must be relevant with regard to the fermion doublers,
so as to give these unwanted species a mass of the order of the cutoff.
Once this is accomplished, one would expect the doublers to decouple in
the continuum theory. This philosophy is reflected in the situation of
QCD with Wilson fermions where the Wilson mass term explicitly breaks
chiral symmetry and accomplishes the desired task of removing the
doublers from the spectrum, giving rise to the standard theory of QCD in
the scaling region. The usual axial anomalies are recovered in the
continuum limit through the effects of the Wilson
term\cite{KarstenSmit}.

With this philosophy in hand, Eichten and Preskill presented a model for
an $SU(5)$ invariant chiral gauge theory. (So as to escape unwanted
gauge
anomalies, they started with a chiral anomaly free $SO(10)$
invariant model and added terms to break the symmetry down to
$SU(5)$.) From strong and weak coupling arguments they discussed
the phase diagram and presented a
scenario that, if realized, would give rise to a theory
of asymptotically free chiral fermions. In the absence
of the generalized Wilson terms, there should be a phase transition
between the strong coupling and weak coupling phases (both symmetric)
across which the fermions become massless. In the strong coupling phase,
right-handed partners to the left-handed fermions form as bound states
of the original fields and these pair up with the left-handed fields to
form massive Dirac fermions. If the
right-handed partners would disassociate when crossing the phase
boundary, the result could be an asymptotically free theory of
massless (but doubled) fermions. When the
generalized Wilson couplings are turned on, the fermion degeneracy would
then be lifted, and the hope was that a wedge would open up in which
only the
right-handed partner of the light fermion would disassociate, with
all others retaining masses of the order of the cutoff.
If a continuum theory is then approached within this wedge in the phase
diagram, a theory of chiral fermions would emerge. A crucial ingredient
in the scenario is that the theory not undergo spontaneous symmetry
breaking.

Although Eichten and Preskill employed generalized Wilson terms in the
form of four point fermion interactions\cite{Ginsparg},
they also pointed out that one
could equivalently formulate the theory in terms of fermions and
scalars, with Yukawa interactions playing the role of the generalized
Wilson terms. In view of the universality hypothesis, both approaches
should produce the same renormalized theory in the scaling region,
although in the former theory the scalar appears as a fermion bound
state rather than as an explicit particle in the lagrangian as is the
case in the Yukawa theory.

Since that original proposal, no further analysis of the model has
appeared in the literature. However, in the mean time several other
methods have been proposed for producing chiral fermions from a lattice
theory [\use{Ref.SmitZak}--\use{Ref.BodwinKovacs}]. In
particular, one of them, the Smit--Swift model
\cite{SmitZak}\cite{Swift}, bears some resemblance to the version of
the Eichten--Preskill proposal in which Yukawa couplings play the role
of generalized Wilson couplings. The Smit--Swift model was first proposed as
a lattice approach to the standard model, and taking its intuition from
the spontaneous symmetry breaking mechanism of the continuum theory,
the Yukawa couplings include generalized Wilson
terms constructed so as to give the original Wilson mass
term in the broken phase of the model at tree level. (Generalized
Wilson terms of this form are usually referred to as
Wilson--Yukawa terms.) Thus the tree level
spectrum in the broken phase is one of massive fermions with the
degeneracy of the doubled fermions lifted by a Wilson mass
term. However, as long as the Wilson--Yukawa coupling is weak, the
doublers do not decouple but merely remain in the low-lying spectrum of
the theory as one tunes to the phase transition, giving rise to a vector
theory\cite{AachenYukawa}.
In the symmetric phase of the weak (Wilson--)Yukawa coupling
region, all fermions become massless and degenerate.

If one tries to remedy the situation by taking the Wilson--Yukawa
coupling strong, the doublers do decouple with masses of the order of
the cutoff, but one pays a penalty. In scaling into the symmetry
breaking phase transition, all interactions vanish, leaving only free
fermions, even in the presence of gauge
fields\cite{JanandUs}. One can modify the model to achieve
non-vanishing gauge interactions, but if one does this, the coupling
turns out to be vector-like. Likewise, deep in the symmetric phase of
the strong Wilson--Yukawa region where one may hope to find
asymptotically free chiral gauge theories, only vector theories
arise\cite{JanandUs}.

The major difference between these Smit--Swift models and the
Eichten--Preskill proposal is in the fermion content. In the former, the
fermions appear in Dirac representations with the goal of
coupling the right- and left-handed components to gauge particles
independently. The Eichten-Preskill model on the other hand contains
only left-handed fermion fields from the outset. Thus
a priori one might hope that the two models would have quite different
behavior and in particular that a chiral theory may still emerge
somewhere in the phase diagram of the Eichten--Preskill model, despite
the apparent similarity of the two models in terms of the structure of the
lagrangian.  If this is not the case, and the phase diagram of the
Eichten--Preskill model is similar to that of the Smit--Swift model,
then existence of the broken phase between the two symmetric phases in
weak and strong Wilson--Yukawa coupling would preclude the possibility
of the scenario originally hoped for by Eichten and Preskill. This is the
question we investigate in the present paper.

The aim of this paper is to present evidence
that the Eichten--Preskill
model does have a phase diagram similar to that of the Smit--Swift model, and
consequently, because of the presence of a symmetry breaking phase
transition, the scenario envisioned by Eichten and Preskill will
most likely not be
realized. We will argue this in several ways.

First of all, to isolate the essence of the problem in view of the fact
that the correct fermion spectrum would have to emerge as a consequence
of the four fermion interactions (or in the Higgs--Yukawa version of the
model, the Yukawa interactions) we turn off the gauge couplings. The
goal of Eichten and Preskill is thus replaced by that of finding a
continuum limit with free, undoubled, left-handed Weyl fermions. Turning
on the gauge fields would then result in an asymptotically free chiral
gauge theory. In order to understand the phase diagram of the model with the
gauge fields turned off, we perform weak and strong coupling expansions.

In the weak
coupling region for both the model with four fermion couplings and the model
with Yukawa couplings, we study the large $N$ solution (where $N$ copies
of the fermion fields are introduced) and match the two models
onto one another to make sure that the expected universality is
satisfied,
at least in this limit.
 In this calculation we explicitly see a symmetry breaking
phase transition which in the model with four fermion coupling should
be interpreted as the Nambu--Jona-Lasinio phase transition.
In strong coupling, we see a similar symmetry breaking
phase transition as in the Smit--Swift model, indicating that there is a
broken phase between the two symmetric phases in the weak and strong
coupling regions respectively.  We present results for the fermion
spectrum across this phase transition, and the fermion spectrum that
we find is always that of Dirac fermions and not that of an undoubled
Weyl fermion.

In the next section we present the two models and discuss some of their
properties. Then we proceed to present our strong coupling arguments,
and subsequently our weak coupling calculations. In our conclusion we
also discuss the coupling of the fermions to external $SO(10)$ gauge
fields. A preliminary report on this work is contained in
\Ref{UsEPJapan}.

%%%#] Introduction:
%%%#[ Models:

\section{$SO(10)$ invariant models\label{sect.models}}

To make contact with phenomenological prejudices at the time, in their
original paper, Eichten and Preskill presented a model that included
symmetry breaking terms to break the $SO(10)$ symmetry down to an
$SU(5)$ symmetry. As this is not necessary to study the essential
features of the proposal, we will investigate models
that keep the $SO(10)$ symmetry intact, so that our target theory will
be an $SO(10)$ invariant asymptotically free chiral gauge theory.
To introduce the models we will study, let us
briefly recall a few elements of group representation theory. Some more
details and a useful representation are
given in Appendix~\use{app.groups}, but for a complete treatment see
e.~g.~\Ref{Georgi}.

First recall that the spinor representations of
$SO(10)$ are obtained as projections from the spinor representation
of $SO(11)$. Thus we introduce a Clifford algebra in $11$
dimensions with generators $\Gamma_a$, $a = 1,\dots,11$.
The fundamental representation for this algebra is
$32$ dimensional and real. In making the projection to $SO(10)$
($a=1,\dots,10$), we use the projection operators
$$
P_{\pm} = {1\over 2}(1\pm \Gamma_{11}),
\EQN{projdef}
$$
which commute with the
generators of $SO(10)$. Thus if we introduce a spinor $\psi$
in the $32$ dimensional spinorial representation of $SO(11)$,
then with respect to
$SO(10)$ it decomposes into two representations distinguished by
the eigenvalues of $\Gamma_{11}$.
The projection operators $P_{\pm}$ project onto the representations
with eigenvalue $+1$ or $-1$ and these representations we denote by $16$
and $\overline{16}$ respectively.
We limit ourselves to Weyl
fermion fields
$\psi$ that transform according to the two-dimensional $(1/2,0)$
(left-handed) representation of $SU(2)\times SU(2)$, and the $16$
representation of $SO(10)$.

Finally, in order to construct lagrangians using these
fields, we introduce the $SO(10)$ invariant tensor $T^{aij}$ defined
through
$$
T^{aij}= \left({C\Gamma_a P_+}\right)^{ij}.
\EQN{invtensor}
$$
where $C$ is the $32$ dimensional charge conjugation
operator (see Appendix~\use{app.groups}).
$T^{aij}$ is a symmetric tensor in the indices $i$ and $j$, transforming
according to the $\overline{16}\times \overline{16}$ representation in
these indices and according to the $10$ representation in $a$. Thus we
can form a scalar bilinear composite field which transforms in
the $10$ representation:
$$
A^a = \psiT \sigma_2 T^a\psi,
\EQN{scalardef}
$$
where $\sigma_2$ is the antisymmetric
Pauli matrix.  This field is in general complex,
with its complex conjugate defined as
$$
A^{*a} = \psidagger \sigma_2 T^{\dagger a}\psi^*.
\EQN{starscalardef}
$$
Note that by virtue of the projection operators, the tensor $T^{\dagger
a}=P_+\Gamma_a C^\dagger$ acts in a different space than $T^a$, with its
indices transforming as a $16\times 16$ representation. There are other
composite scalar fields, bilinear in the fermions, which transform in the
$126$ and $\overline{126}$ representations, but these turn out not to
propagate, so we will not consider them in this paper.

The scalar fields $A$ and $A^*$ defined in Eqs.~\use{Eq.scalardef}
and \use{Eq.starscalardef}
are interchanged by $CP$ symmetry:
$$
\EQNalign{
\psi(t,{\vec x}) &\to C\sigma_2
\psi^*(t,-{\vec x}), \EQN{chargeconj}\cr
\Dmu &\to (\D_0,-\D_i). \EQN{parity}\cr
}
$$
Hence, if we define
$$
A^a = {1\over\sqrt2}(A^a_1 + iA^a_2), \EQN{scalarseparate}
$$
the real and imaginary parts are even and odd respectively under $CP$
transformations:
$$
\EQNalign{
A^a_1(t,{\vec x})&\to A^a_1(t,-{\vec x}), \EQN{Achargeconj} \cr
A^a_2(t,{\vec x})&\to-A^a_2(t,-{\vec x}). \cr
}
$$

We can now write down the action of the $SO(10)$ invariant version of the
Eichten--Preskill model (with the gauge interaction turned off):
$$
\SEP = \sum_x\lbrack\Lagkin + \Lagfour + \LaggenWilson\rbrack,
\EQN{fourfermilag}
$$
where the kinetic term for the fermions is
$$
\Lagkin =
\half\sum_{\pm\mu}\epsilon(\mu)\psidagger(x)\sigmahat_\mu\psi(x+\muhat),
\EQN{fermikinlag}
$$
in which
$$
\EQNalign{
\sigmahat &= (1,i\vec\sigma), \EQN{sigmahat}\cr
\epsilon(\mu) &= \hbox{\rm sign}(\mu) = \pm1,\cr
}
$$
where $\vec\sigma = (\sigma_1,\sigma_2,\sigma_3)$ are the Pauli
matrices. The four point interaction term is
$$
\Lagfour =
- {\lambda\over24}
\sum_{a}\left(\psiT(x)\sigma_2 T^a\psi(x)\right)^2+\hc,
\EQN{Lagfour}
$$
and the generalized Wilson term is given by
$$
\LaggenWilson = - {r\over48}
\sum_{a}\Delta\left(\psiT(x)\sigma_2T^a\psi(x)\right)^2+\hc,
\EQN{laggenWilson}
$$
where
$$
\EQNalign{
\Delta(\psi_i\psi_j\psi_k\psi_l) = -{1\over 2}\sum_{\pm\mu}
 & \lbrack\psi_i(x+\muhat)\psi_j(x)\psi_k(x)\psi_l(x)
 \EQN{defDelta}\cr
 & +\psi_i(x)\psi_j(x+\muhat)\psi_k(x)\psi_l(x)\cr
 & +\cdots -4\psi_i(x)\psi_j(x)\psi_k(x)\psi_l(x)\;\rbrack.\cr
}
$$
Note that these interaction terms are constructed to explicitly break the
left-handed fermion number (cf. baryon
number in the standard model)
which is anomalous in the continuum theory, so the model does not
suffer from the problem of an unwanted global invariance as recently raised by
T.~Banks[\use{Ref.Banks}--\use{Ref.Banksnew}].

The quantum theory is now defined through the path integral
$$
\langle\cdots\rangle =
Z^{-1}\int\dfermi e^{-\SEP}(\cdots),
\EQN{EPev}
$$
where $Z$ is defined so that $\langle1\rangle=1$.
As already
pointed out by Eichten and Preskill, a model with the same symmetries
and particle content can be constructed through a lagrangian that
couples scalars in the $10$ representation to the fermions
through Yukawa couplings. Formally, if one sets $r=0$, an identification
can be made through the one point integral ($y^2/b=\lambda/6$)
$$
e^{{\lambda\over24}
\sum_{a}\left(\psiT(x)\sigma_2 T^a\psi(x)\right)^2+\hc}=
{
{
\int\lbrack d\phi_1\rbrack\lbrack d\phi_2\rbrack e^{-\half b\phi_1^2 -
\half b\phi_2^2
-\half y
\sum_a\lbrack\phi_1^a(A^a+A^{*a})+\phi_2^a(A^a-A^{*a})\rbrack}
}
\over
{
\int\lbrack d\phi_1\rbrack\lbrack d\phi_2\rbrack e^{-\half b\phi_1^2 -
\half b\phi_2^2}
}
},
\EQN{onepoint}
$$
where $A$ and $A^*$ have been defined in Eqs.~\use{Eq.scalardef}
and \use{Eq.starscalardef}.
Note that the field $\phi_2$ couples to an anti-hermitian fermion
bilinear.  In order to later compare correlation functions computed in the
Yukawa version of the model  with those computed in the four fermion
version, one therefore has to analytically continue $\phi_2\to i\phi_2$.
We will return to this point later.

Thus we may omit the four fermion interaction in favor of introducing
scalar fields coupled to the fermions via the Yukawa couplings:
$$
\LagY= \half y \sum_a\left(\phi_1^a\lbrack\psiT\sigma_2
T^a\psi+\hc\rbrack
+\phi_2^a\lbrack\psiT\sigma_2 T^a\psi-\hc\rbrack\right),
\EQN{lagyukawa}
$$
and an integration over the scalar fields.
In general we should also include kinetic terms for the scalar fields:
$$
\EQNalign{
\Lagscalarkin =&-\sum_{a,\mu}
\lbrack\kappa_1\phi_1^a(x)\phi_1^a(x+\muhat)+
\kappa_2\phi_2^a(x)\phi_2^a(x+\muhat)\rbrack\cr
&+\half
b\sum_a\lbrack\phi_1^a(x)\phi_1^a(x)+\phi_2^a(x)\phi_2^a(x)\rbrack.
\EQN{lagscalarkin}}
$$
Finally we need to add a generalized Wilson term to play the role of
$\LaggenWilson$ above. For this we choose a Wilson--Yukawa term,
$$
\LagWY= - \quart
w\sum_a\left(\phi_1^a\lbrack\psiT\sigma_2T^a\Box\psi+\hc\rbrack
+\phi_2^a\lbrack\psiT\sigma_2 T^a\Box\psi-\hc\rbrack\right),
\EQN{WilsonYukawa}
$$
where $\Box$ is the lattice laplacian:
$$
\EQNalign{
\Box \psi(x) &= \sum_\mu\Dmu\Dmutilde\psi(x), \EQN{boxdef}\cr
\Dmu\psi(x) &= \psi(x+\muhat)-\psi(x),\cr
\Dmutilde\psi(x) &= \psi(x)-\psi(x-\muhat).\cr
}
$$
Thus the full lagrangian for the Eichten--Preskill model with Yukawa
couplings is:
$$
\LagEPY = \Lagkin + \Lagscalarkin + \LagY+\LagWY.
\EQN{lagEPY}
$$
The path integral again
defines the theory as in \Eq{EPev}, but now
$SO(10)$ invariant integrals over the scalar
fields must also be performed.

Both models are explicitly $SO(10)$ chiral invariant, and this $SO(10)$
symmetry can be gauged straightforwardly.  Both models are also
invariant under $CP$ symmetry, \Eq{chargeconj} and \Eq{Achargeconj},
with $\phi_1$ and $\phi_2$ transforming like $A_1$ and $A_2$
respectively.

%%%#] Models:
%%%#[ Strong coupling:

\section{Strong coupling calculations in the Eichten--Preskill
model\label{sect.SCinEP}}

In this section we begin our analysis by repeating
the strong coupling arguments given by Eichten and Preskill, for our
specific model. Thus we start with the lagrangian given by
\Eq{fourfermilag}, and first treat the case of $r=0$. In order to
facilitate the expansion, we rescale $\psi$:
$$
\psi\to{1\over{\lambda^{1/4}}}\psi,
\EQN{SCrescalepsi}
$$
so that the lagrangian becomes $(r=0)$
$$
\LagEP =
- {1\over24}
\sum_{a}\left\lbrack(\psiT(x)\sigma_2 T^a\psi(x))^2+\hc\right\rbrack
+ \half\alpha\sum_{\pm\mu}
\epsilon(\mu)\psidagger(x)\sigmahat_\mu\psi(x+\muhat),
\EQN{SClagEP}
$$
where
$$
\alpha={1\over\sqrt\lambda}.
\EQN{alpha}
$$
Using the four point
fermion term as the zeroth order lagrangian in the path integral, we may
expand the expectation value of any combination of fields in terms of
$\alpha$ to first non-trivial order by using the one point integral
$$
\EQNalign{
\langle\psi_{i\alpha}(x)&\psi_{j\beta}(x)\psi_{k\gamma}(x)
\psi_{l\delta}(x)\rangle_0 \EQN{onepointintegral}\cr
&= Z_0^{-1}\int\dfermi e^{-\sum_z\Lagfour(z)}
\psi_{i\alpha}(x)\psi_{j\beta}(x)\psi_{k\gamma}(x)\psi_{l\delta}(x)\cr
&={1\over 80}\left\lbrack
 \sigma_{2\alpha\beta}\sigma_{2\gamma\delta}T^{\dagger c}_{ij}T^{\dagger
c}_{kl}
-\sigma_{2\alpha\gamma}\sigma_{2\beta\delta}T^{\dagger c}_{ik}T^{\dagger
c}_{jl}
+\sigma_{2\alpha\delta}\sigma_{2\beta\gamma}T^{\dagger c}_{il}T^{\dagger
c}_{jk}
\right\rbrack,\cr
}
$$
where a sum over $c$ is implied and the zero subscript on the
expectation value and on $Z_0$ stands for $\alpha=r=0$.
This formula can be used to find a recursion relation for the fermion
propagator in the strong coupling region, by
expanding the quantity
$$
\langle\psi_{i\alpha}(x)\psi^{*j}_{\beta}(0)\rangle =
{\int\dfermi e^{-\sum_z \LagEP(z)}
\psi_{i\alpha}(x)\psi^{*j}_{\beta}(0)\over
\int\dfermi e^{-\sum_z \LagEP(z)}},
$$
in terms of $\alpha$ and integrating over the fermion fields at site $x$.
In the present case,
the first non-vanishing contribution to
the expansion comes from the $\alpha^3$ term, which gives rise to the
relation
$$
\langle\psi_{i\alpha}(x)\psi^{*j}_{\beta}(0)\rangle =
({\alpha\over 2})^3\sum_{\pm\mu,\alpha^\prime}
\epsilon(\mu)\sigmabar_{\mu,\alpha\alpha^\prime}\langle
\B_{i\alpha^\prime}(x+\muhat)\psi^{*j}_{\beta}(0)\rangle
+O(\alpha^7),
\EQN{recpart1}
$$
where
$$
\sigmabar_\mu = (1, -i\vec\sigma),
$$
and $\B$ is a composite ``baryon'' operator with the quantum numbers of a
right-handed fermionic partner to $\psi$:
$$
B = {1\over 160}\sum_a\sigma_2
T^{\dagger a}\psi^*(\psidagger\sigma_2T^{\dagger a}\psi^*).
\EQN{Bdef}
$$
Similarly, one can derive the related relation
$$
\langle\B_{i\alpha}(x)\psi^{*j}_{\beta}(0)\rangle =
-{3\over 80}\left\lbrack\delta_{i}^{j}\delta_{\alpha\beta}\delta_{x0}
-{\alpha\over2}\sum_{\pm\mu,\alpha^\prime}\epsilon(\mu)\sigmahat_{\mu,
\alpha\alpha^\prime}
\langle\psi_{i\alpha^\prime}(x+\muhat)\psi^{*j}_{\beta}(0)\rangle
\right\rbrack(1+O(\alpha^4)),
\EQN{recpart2}
$$
which when combined with \Eq{recpart1} leads to the
propagator for the field $\psi$,
$$
\langle\psi\psidagger\rangle(p) =
{-{1\over\alpha}\sum_\mu\sigmabar_\mu\;i\sin p_\mu\over
\sum_\mu \sin^2p_\mu + {320\over{3\alpha^4}}}
+O(\alpha^7)
$$
in momentum space.   From this it follows that
in the strong coupling region the fermion is massive
with a mass
$$
m_{\rm fermion}=8\sqrt{{5\over 3}}\lambda+O({1\over\lambda}).
\EQN{fermionmass}
$$
As was also
shown by Eichten and Preskill, the first correction to this for non-zero
$r$ lifts the degeneracy of the doubled fermions. Thus, to recount their
conclusions, in the strong $\lambda$ region of the phase diagram
a three fermion bound state appears with the quantum numbers of a
right-handed partner for $\psi$ and together they form a massive Dirac fermion.

A similar calculation can be performed for the propagator of the scalar
field defined in \Eq{scalardef}.  Starting with
$$
\langle A^a(x)A^b(0)\rangle = Z^{-1}\sum_{n=0}^\infty{1\over n!}
\int\dfermi e^{-\sum_z\Lagfour(z)}(-\sum_z\Lagkin(z))^n A^a(x)A^b(0),
\EQN{bospropexpans}
$$
and applying
\Eq{onepointintegral}, one finds to lowest non-trivial order
$$
\EQNalign{
\langle A^{a}(x)A^b(0)\rangle = &{96\over 5}\delta_{ab}\delta_{x0}+
({\alpha\over 2})^2{3\over 10}\sum_{\pm\mu}
\langle A^{*a}(x+\muhat)A^b(0)\rangle, \EQN{bosprop}\cr
\langle A^{*a}(x)A^b(0)\rangle =
&({\alpha\over 2})^2{3\over 10}\sum_{\pm\mu}
\langle A^{a}(x+\muhat)A^b(0)\rangle. \EQN{bosprop}\cr
}
$$
Separating the scalar field into its real and imaginary parts as in
\Eq{scalarseparate}, and redefining
$$
A_2(x) \to (-1)^{\scriptscriptstyle \sum_\mu x_\mu}A_2(x),
\EQN{staggeredA2}
$$
which corresponds to a shift $p\to p+(\pi,\pi,\pi,\pi)$ in momentum
space, the propagators for these components are
$$
\EQNalign{
D_1^{ab}(p) &= {96\over 5}
\delta^{ab}(1 - {3\over 20}\alpha^2\sum_\mu\cos p_\mu)^{-1},
\EQN{bospropp}\cr
D_2^{ab}(p) &= -{96\over 5}
\delta^{ab}(1 - {3\over 20}\alpha^2\sum_\mu\cos p_\mu)^{-1}.\cr
}
$$
Thus it turns out that there are two degenerate scalars (to this order),
both with mass
$$
\EQNalign{
m^2 &= 8\lambda\left({5\over 3}-{1\over\lambda}+
O({1\over{\lambda^2}})\right).\EQN{bosonmasses}\cr
}
$$
However, the propagator for $A_2$ has a negative residue and is
therefore unphysical.

This result suggests that there must be an
$SO(10)\to SO(9)$  symmetry breaking phase
transition for some finite positive value of $\lambda$.
The $A_1$ scalar field goes from being a massive scalar field
at strong positive coupling to a field of negative mass squared for
$\lambda$ smaller than a critical value, which from \Eq{bosonmasses}
is $\lambda_c=0.6$ to lowest order.  A disease of this model is that at
this critical value also the unphysical $A_2$ scalar field becomes
massless.  This disease can however be cured by the adjustment of
additional parameters, as will become clear in the next section, where
we will make contact between this model and the model with Yukawa
couplings.

\section{Strong coupling calculations in the Yukawa model\label{sect.SCinY}}

Now let us perform strong coupling calculations for the model
with Yukawa couplings defined through the lagrangian given in\Eq{lagEPY}.
We will employ a large $N$ technique suitable for large values of the
Yukawa couplings \cite{Kutietal}.  The fermions are given an extra index
$n$
which is summed over from $1$ to $N$.  This introduces an extra $SO(N)$
symmetry into the model.  Notice that no anomalies are introduced
through this procedure.  In this section we will also rescale some
couplings with $N$ as follows:
$$
y=\ytilde\sqrt{N},\ \ \ \
w=\wtilde\sqrt{N},\ \ \ \ \
b={\tilde b} N,\EQN{rescaling}
$$
and keep $\ytilde$, $\wtilde$, $\tilde b$, $\kappa_{1}$ and
$\kappa_{2}$ fixed.
With these alterations, the lagrangian now is
$$
\EQNalign{
\LagEPY =
&\half{\tilde b}N\sum_a\lbrack\phi_1^a(x)\phi_1^a(x)
+\phi_2^a(x)\phi_2^a(x)\rbrack \cr
-&\kappa_1\sum_{a,\mu} \phi_1^a(x)\phi_1^a(x+\muhat)
-\kappa_2\sum_{a,\mu} \phi_2^a(x)\phi_2^a(x+\muhat)\cr
+&\half\ytilde\sqrt{N}
\sum_{a,n}\left(\lbrack\psiT_n(x)\sigma_2 T^a\psi_n(x)+\hc\rbrack
\phi_1^a(x)
+\lbrack\psiT_n(x)\sigma_2 T^a\psi_n(x)-\hc\rbrack\phi_2^a(x)\right)\cr
- &\quart\wtilde\sqrt{N} \sum_{a,n}
\left(\lbrack\psiT_n(x)\sigma_2T^a\Box\psi_n(x)+\hc\rbrack\phi_1^a(x)
+\lbrack\psiT_n(x)\sigma_2T^a\Box\psi_n(x)-\hc\rbrack\phi_2^a(x)\right)\cr
+&\half\sum_{\pm\mu,n}\epsilon(\mu)\psidagger_n(x)
\sigmahat_\mu\psi_n(x+\muhat).
\EQN{LlargeN}\cr
}
$$
Since the fermionic part of the lagrangian,
$\left(\LagEPY\right)^{\rm fermion}$
is bilinear in the fermion fields, we may define a
matrix ${\cal S}_{IJ}$ through
$$
S^{\rm fermion}=\sum_x\left(\LagEPY\right)^{\rm fermion}(x)=
\half\sum_{IJ}\Psi_I {\cal S}_{IJ}\Psi_J,
\EQN{Sdef}
$$
where $I$ is a composite index running over $x$, the Weyl spinor index
$\alpha$, the $SO(10)$ index $i$, the $SO(N)$ index $n$ and the
`components' $\psi$ and $\psi^*$ of the fermion field $\Psi$.  By
integrating out the fermions we obtain an effective action for the
fields $\phi_1$ and $\phi_2$
$$
\EQNalign{
S_{\rm eff}=
+&\half{\tilde b}N\sum_{x,a}\lbrack\phi_1^a(x)\phi_1^a(x)
+\phi_2^a(x)\phi_2^a(x)\rbrack\cr
-&\kappa_1\sum_{x,a,\mu} \phi_1^a(x)\phi_1^a(x+\muhat)
-\kappa_2\sum_{x,a,\mu} \phi_2^a(x)\phi_2^a(x+\muhat)\cr
-&\half\log\det{\cal S}.              \EQN{Seffdef}\cr
}
$$
To $O(1/N)$, $S_{\rm eff}$ is
$$
\EQNalign{
S_{\rm eff}=
&\half{\tilde b}N\sum_{x,a}\lbrack\phi_1^a(x)\phi_1^a(x)
+\phi_2^a(x)\phi_2^a(x)\rbrack \EQN{Seff}\cr
&-8N\sum_x\lbrack\log{\phi_+^2(x)}+\log{\phi_-^{2}(x)}\rbrack\cr
&-2{\tilde\beta}^2 J({\tilde\beta}\wtilde)
\sum_{x,\mu,a}
\left({{\phi_-^{a}(x+\muhat)\phi_+^a(x)}\over{\phi_-^{2}(x+\muhat)\phi_+^2(x)}}
+\hc\right)\cr
&-\kappa_1\sum_{x,a,\mu} \phi_1^a(x)\phi_1^a(x+\muhat)
-\kappa_2\sum_{x,a,\mu} \phi_2^a(x)\phi_2^a(x+\muhat),\cr
}
$$
where
$$
\phi_\pm(x)=\phi_1(x)\pm\phi_2(x),\ \ \ \ \
{\tilde\beta}={1\over{4\wtilde+\ytilde}}\EQN{somedefs}
$$
and
$$
J(z)=
\int_p{{\sum_\mu\sin^2{p_\mu}}\over{(1-z\sum_\mu\cos{p_\mu})^2}},
\EQN{Jdef}
$$
in which
$$
\int_p\equiv{1\over{(2\pi)^4}}\int\int\int\int_{-\pi}^{\pi}d^4p.\EQN{intp}
$$
Finally, for $N\to\infty$ we may employ the saddlepoint approximation to
find
$$
\phi_\pm^2=\sum_a\phi_\pm^a\phi_\pm^a={32\over{\tilde b}}.\EQN{freeze}
$$
Rescaling the boson fields such that $\phi_\pm^2=1$
and redefining (cf. \Eq{staggeredA2})
$$
\phi_2(x)\to (-1)^{\scriptscriptstyle \sum_\mu x_\mu}\phi_2(x) \EQN{phi2redef}
$$
 leads to a simple
form for the effective action in the large $N$ limit:
$$
\EQNalign{
S_{\rm eff}=
&-{1\over 8}{\tilde b}{\tilde\beta}^2 J({\tilde\beta}\wtilde)
\sum_{x,\mu,a}\lbrack\phi_1^a(x)\phi_1^a(x+\muhat)
+\phi_2^a(x)\phi_2^a(x+\muhat)\rbrack\cr
&-\kappa_1\sum_{x,a,\mu} \phi_1^a(x)\phi_1^a(x+\muhat)
+\kappa_2\sum_{x,a,\mu} \phi_2^a(x)\phi_2^a(x+\muhat),\EQN{Sefffinal}
}
$$
with constraints
$$
\sum_a[(\phi_1^a)^2+(\phi_2^a)^2]=1,\ \ \ \ \ \sum_a\phi_1^a\phi_2^a=0.
\EQN{constraints}
$$
We now have to remember that $\phi_2$ should be
continued to $i\phi_2$ (cf. \Eq{onepoint}), which means that
the model contains
unphysical states (for $\kappa_1=\kappa_2=0)$,
as we found already in the previous section.  The
model with lagrangian \Eq{SClagEP} corresponds to the parameter choice
$\kappa_1=\kappa_2=0$ and $w=0$ (hence $J(w=0)=2$).  The critical value
for $\lambda$ which we found in that model in strong coupling,
$\lambda_c=0.6$,
leads to a critical value for the effective hopping parameter
$$
(\kappa_{\rm eff})_c\equiv({{\tilde b}\over{4\ytilde^2}})_c
={3\over{2\lambda_c}}={5\over 2}.\EQN{keff}
$$
This corresponds to the mean field value for the critical hopping
parameter for the effective action in \Eq{Sefffinal} with
$\kappa_1=\kappa_2=0$.
Note that $\lambda=6y^2/b$ is independent of $N$.

It is
now also clear how the problem of the unphysical $A_2$ field
can be solved: by adjusting $\kappa_2$
and/or adding a scalar potential for the field $\phi_2$ one can arrange
that even in those regions of the phase diagram where the $\phi_1$ mass
becomes physical (i.e. becomes small in lattice units), the $\phi_2$
mass remains of the order of the cutoff ($O(1)$ in lattice units).
In general however, the model is not unitary (cf. also
\cite{Jantransfer}).  These additional parameters would presumably
correspond to higher derivative terms in the four fermion formulation
of the model \cite{Kutietal}.

Next we come to the fermion two-point function.
%The scalar propagatorsappear in the expansion of course,
%and they subsequently must be
%approximated with whatever method is appropriate for a particular
%place in the phase diagram. For example, near the critical
%point we can use mean field theory for these propagators, whereas deep
%in thesymmetric phase we can expand in small $\kappa$.
%Here we will not give a complete review of how the expansion is
%performed, but only point out relevant differences from the case of the
%Smit--Swift model studied previously\cite{Usoneoverw}. The main
%difference is that in the case of the Smit--Swift model, one could make
%a change of variables to a fermion field which did not transform under
%the group to be gauged, the so-called `neutral' field in \Ref{Usoneoverw}.
%Such a transformation is not possible in the Eichten--Preskill model
%because of the structure of the lagrangian.
%This changes the diagrammatics of the expansion,
% but the expansion can nevertheless be performed because of the
%properties of the matrices $T^a$. In particular, if we define the
%field $\Phi(x)$
First, we need to define an operator
in this model analogous to $\B$ of the four fermion model given in
\Eq{Bdef} to excite right-handed fermion states. From \Eq{scalardef} one
sees that a composite operator which transforms as a right-handed
fermion can be defined by
$$
\B(x) = \sum_a\phi^{*a}(x)T^{\dagger a}\sigma_2\psistar(x).
\EQN{Bdef2}
$$
The result for the fermion propagator $S(p)$ in momentum space is then
$$
S(p) =
{{-i\sum_\mu\gamma_\mu\sin{p_\mu}[P_R
+P_L(z^2+O(1/N))]+M(p)}\over {M^2(p)+\left(z^2+
O\left(1/N\right)\right)\sum_\mu\sin^2{p_\mu}}}
\left(1+O(1/N^2)\right),\EQN{fermtwopoint}
$$
where
$$
M(p)=y+w\sum_\mu(1-\cos{p_\mu}),\EQN{Mdef}
$$
and
$$
z^2={{\tilde b}\over{32}}\sum_a\left\langle\phi^a_1(x)\phi^a_1(x\pm\muhat)
+\phi^a_2(x)\phi^a_2(x\pm\muhat)\right\rangle,\EQN{zsqdef}
$$
where the fields $\phi_{1}$ and $\phi_{2}$
correspond to those in \Eq{Sefffinal}.
Thus as in the case of the Smit--Swift model, we find one massless Dirac
fermion and the doublers have masses of the order of the cutoff (for
$y\to 0, w$ finite).

The large $N$ technique used in this section is very similar to the
$1/w$ expansion developed in \Ref{Usoneoverw}. As in that case, the
approach allows us to integrate out the fermions systematically through
perturbation theory in a small parameter (here $1/N$), but it does not
provide us with a way of computing the bosonic correlation functions
that remain, as for instance \Eq{zsqdef}. To this order in $1/N$, this
is however the only bosonic correlation function we need, and $z^2$ is
finite and non-zero at the phase
transition\cite{Usoneoverd}\cite{Aachenoneoverw}. Notice that this
``incompleteness'' prevents us from employing this technique at strong
coupling in the four fermion formulation of the model, as is clear from
the fact that $\lambda$, defined through \Eq{onepoint}, is independent
of $N$.

Finally we summarize what we have learned from
strong coupling calculations in the Yukawa model. First the phase
diagram looks very similar to that of the Smit--Swift model, with a pure
scalar theory at $\ytilde=\infty$ and/or
$\wtilde=\infty$ (${\tilde\beta}^2 J({\tilde\beta}\wtilde)=0$
at $\ytilde=\infty$ and/or
$\wtilde=\infty$, cf.
\Eq{Sefffinal}) that undergoes symmetry breaking at some
critical value of $\kappa_{1,2}$.
As the fermions are turned on with $\ytilde$ and $\wtilde$ finite,
this symmetry breaking transition moves down in $\kappa_{1,2}$.
If we set $w=0$, $\kappa_{1}=0$ and $\kappa_{2}=0$
this transition point should correspond to the
phase transition point in the four fermion model where $\lambda =
\lambda_c$.  The fact that the value found for $\lambda_c$ in the
previous section corresponds to the mean field critical point for the
effective action \Eq{Sefffinal}
lends credence to the conclusion that the four
fermion model also undergoes a symmetry breaking phase transition, and
that in fact the estimate $\lambda_c=0.6$ is reasonably reliable.
{}From \Eq{fermtwopoint} we infer that the right-handed fermion does not
disassociate over the phase transition
and hence, within the range of validity
of our expansion, the scenario envisaged in \Ref{EichPres} does not occur.

%%%#] Strong coupling:
%%%#[ Large N:
\section{Large N calculations at weak coupling\label{sect.largeN}}

In this section we turn to a study of the two models
using the weak coupling
large $N$ expansion.
Again, the fermions are given an extra index which is
summed over from $1$ to $N$,
but now the couplings are scaled with $N$ in a way appropriate for the
weak coupling domain.
In this case, both models can be solved exactly to leading
order in $1/N$, and thus a direct comparison can be performed between
the two models. A similar calculation for a comparison between the
standard model and the top quark condensate model has been performed in
\Ref{Kutietal}. A nice review of large $N$
techniques is given in \Ref{Coleman}.

The weak coupling
$1/N$ expansion for the four fermion model is an expansion
in $1/N$ keeping
$$\ltilde = N\lambda \EQN{lambdatildedef}$$
and
$$\rtilde=Nr \EQN{rtildedef}$$
fixed, which
means that the expansion is valid in the region where $\lambda$ and $r$
are small. Thus we will be probing the perturbative region of the theory
using this expansion, and the doublers will be present. Nevertheless, we
can still make the desired comparisons to gain some
confidence that the two models we are dealing with do indeed result in
the same physics in the scaling region. Moreover, this will lead to
independent confirmation that a broken phase exists between the
symmetric phases at weak and strong coupling. Since the techniques are
well known, we will only present our results here.

As a first result, we calculate the mass gap for the fermion field, within
the four fermion model. This mass gap serves as an order parameter for
symmetry breaking, because $SO(10)$ symmetry forbids such a term
in the lagrangian. After the usual expansion in Feynman diagrams, keeping
only leading terms in $1/N$ we obtain
an equation for the inverse fermion propagator,
$$
S^{-1}(p) = \lbrack\langle\Psi\Psidagger\rangle(p)\rbrack^{-1} =
S_0^{-1}(p) + \Itilde(p), \EQN{Nfermprop}
$$
where we have written
$$
\Psi = \pmatrix{\psi\cr\sigma_2\psi^*}, \EQN{Npsidef}
$$
and $S_0^{-1}$ and $\Itilde(p)$ are matrices
in the space of upper and lower, or
left-handed and right-handed components of the fermions.
$S_0$ is the free propagator:
$$
S^{-1}_0(p) = \pmatrix{i\sum_\mu \sigmahat_\mu \sin p_\mu&0\cr
0&i\sum_\mu \sigmabar_\mu \sin p_\mu\cr}, \EQN{S0def}
$$
and $\Itilde$ is equal
to the amputated
one loop tadpole diagram containing the exact resummed propagator
in keeping with the $1/N$ expansion, closed by a four fermion
vertex. If we parametrize the propagator by
$$
S^{-1}(p) = \pmatrix{i\sum_\mu \sigmahat_\mu \sin p_\mu&
\sum_a\eta^a T^\dagger_a \Sigma(p)\cr
\sum_a\eta^a T_a \Sigma(p)&i\sum_\mu \sigmabar_\mu \sin p_\mu\cr},
\EQN{propparam}
$$
where $\eta^a$ is a unit vector in the ten dimensional space indicating
the direction of possible symmetry breaking, we can write
$$
\Itilde(p) = \pmatrix{0&\Idagger(p)\unity\cr I(p)\unity&0\cr}, \EQN{Imatdef}
$$
in which
$$
I(p) = {16\over 3} \sum_a\eta^a T_a \int_q
{\Sigma(q)\over D(q)}
\lbrack\ltilde + \rtilde\sum_\mu (2-\cos p_\mu -\cos q_\mu)\rbrack,
\EQN{Idef}
$$
where
$$
D(q) = \sum_\mu \sin^2q_\mu + \Sigma^2(q).
\EQN{Ddef}
$$
This result is exact in the limit that $N\to\infty$, and leads to the
gap
equation for the self energy $\Sigma(p)$:
$$
\Sigma(p) = {16\over 3}\int_q{\Sigma(q)\over
D(q)}\lbrack \ltilde
+\rtilde\sum_\mu (2-\cos p_\mu -\cos q_\mu)\rbrack.
\EQN{Sigmadef}
$$
This can be rewritten
$$
\Sigma(p) =  f\rtilde C(p) + \Sigma(0),
\EQN{Sigmaparameterization}
$$
where
$$
 f = {16\over 3}\int_q{\Sigma(q) \over D(q)}
\EQN{fdef}
$$
is just a constant (dependent on the couplings), and
$$
C(p) = \sum_\mu(1-\cos p_\mu)
\EQN{Cdef}
$$
isolates all the momentum dependence in the first term of
\Eq{Sigmaparameterization} and vanishes for $p=0$.
For $\Sigma(p)$ to vanish, clearly both terms on the right hand side of
\Eq{Sigmaparameterization} must vanish independently. The converse is
also true: for $\Sigma$ to be non-vanishing, neither $\Sigma(0)$ nor
$ f$ will vanish. Thus we can write
$$
\Sigma(0) =  f\rho,
\EQN{rhodef}
$$
where $\rho$
is another constant dependent on the couplings.
With this parametrization, the symmetric phase for
which $\Sigma$ vanishes is characterized by a vanishing value of
$ f$.

To look for critical values of the couplings between a
symmetric and a broken phase, we reinsert the parametrization of
Eqs.~\use{Eq.Sigmaparameterization} and \use{Eq.rhodef} into
\Eq{Sigmadef}. Equating this to \Eq{Sigmaparameterization} gives us
two equations:
$$
\EQNalign{
\rho &= \rtilde^2 \I2 + \rtilde(\ltilde+\rho)\I1+\ltilde\rho \I0, \EQN{A1}\cr
1&=\rtilde \I1+\rho \I0,\cr
}
$$
where $\I{n}$ is the integral
$$
\I{n} = {16\over 3}\int_q {C(q)^n\over\sum_\mu \sin^2q_\mu +
 f^2(\rtilde C(q)+\rho)^2},
\EQN{Intdef}
$$
for $n=0,1,2$.  Solving for $\rho$ in the second equation and putting that back
into the first finally gives us the equation
$$
\ltilde = {(\I1^2-\I2 \I0)\rtilde^2 - 2\rtilde \I1 + 1\over \I0}.
\EQN{criticalsurface}
$$
All three
integrals $\I0$, $\I1$ and $\I2$ are positive, and furthermore because
$$
\ltilde^2 \I0 + 2\rtilde\ltilde \I1 + \rtilde^2 \I2 =
{16\over 3}\int_q{(\ltilde+\rtilde C(q))^2\over D(q)}
> 0,
\EQN{Iinequality1}
$$
we have
$$
\I0 \I2 > \I1^2.
\EQN{Iinequality2}
$$
As $ f\to 0$, the integrals defined in \Eq{Intdef} become pure numbers
independent of the couplings, and in fact
$$
\EQNalign{
I_1( f=0)&=4I_0( f=0), \EQN{Irelations}\cr
I_2( f=0)&=20I_0( f=0)-{16\over 3}.\cr
}
$$
\Eq{criticalsurface} becomes an
equation for the
critical coupling $\ltilde^*$ as a function of $\rtilde$
and \Eq{Iinequality2} tells us that
this represents an inverted parabola for $\ltilde$
as a function of $\rtilde$.
We thus clearly see that for all values of $\rtilde$, the large $N$
solution has a broken phase, for
$\ltilde$ larger than some critical value $\ltilde^*$.  For $\rtilde=0$
we find
$$
\ltilde^*={1\over{I_0( f=0)}}=0.3025\;.\EQN{lambdastar}
$$
This value is sufficiently smaller, even at $N=1$,
than the value at which the strong
coupling phase transition occurs to infer
that a broken phase does indeed separate the symmetric phases at weak
and strong coupling.

A similar calculation can be done for the bosons in this model as
represented by the composite operators defined in Eqs.~\use{Eq.scalardef}
and \use{Eq.starscalardef}. This propagator in the $1/N$ expansion is
given by a summation of all diagrams consisting of simple loops joined
by four point vertices.  The calculation is again straightforward,
so we will omit the details and simply present
the results. Assuming a symmetry breaking to take place in the
$1$ direction in the $10$ dimensional space of $SO(10)$
we find the following result for the propagators. The propagator for the
Higgs field, $A^1_1$, in momentum space is
$$
\EQNalign{
\langle A^1_1 A_1^1\rangle(p) &=
{{2NH^-_0(p)}\over 1-{\ltilde\over 6}{\cal H}^-(p)}, \EQN{largeNHiggsprop}\cr
\noalign{\hbox{and that for the Goldstone modes is}}
\langle A^j_1 A^j_1\rangle(p) &=
{{2NH^+_0(p)}\over 1-{\ltilde\over 6}{\cal H}^+(p)},
\;\;j=2,\ldots,10, \EQN{largeNGoldstoneprop}\cr
}
$$
where
$$
{\cal H}^\pm(p) = \H0\pm(p)+2{\rtilde\over\ltilde}\H1\pm(p)+
{\ltilde\over
6}({\rtilde\over\ltilde})^2(\H0\pm(p)\H2\pm(p)-\H1\pm(p)^2),
$$
in which
$$
\H{n}\pm(p) = 32\int_q\lbrack\half(C(q)+C(p-q))\rbrack^n
{
\sum_\mu\sin q_\mu\sin(q_\mu-p_\mu)\pm\Sigma(q)\Sigma(q-p)
\over
D(q) D(q-p)
},
\EQN{Hdef}
$$
and $C(q)$, $D(q)$ and $\Sigma(p)$ are as defined in Eqs.~\use{Eq.Cdef},
\use{Eq.Ddef} and \use{Eq.Sigmaparameterization} respectively. It should be
noted that the Higgs field has a mass
proportional to $\Sigma$ and the mass of the Goldstone field vanishes
identically. For vanishing $\Sigma$, the two propagators are identical
as expected and no symmetry breaking occurs.

Finally we give the result of the effective Yukawa coupling for this
model (the one particle irreducible three point function) which is
$$
\EQNalign{
\langle A_1^a(p)\Psi(q)\Psidagger(-p-q)\rangle_{\rm 1PI}
&\equiv\pmatrix{0&T^{\dagger a}\cr T^a&0\cr}
y_{\scriptscriptstyle\rm EP}(p,q),
\EQN{effectiveYukawacoupling1}\cr
y_{\scriptscriptstyle\rm EP}(p,q)&={\Sigma(q)+\Sigma(p+q)\over 2v},
}
$$
where
$$
v = \langle A^1_1\rangle = -{6\sqrt2N f},
\EQN{vvalinEP}
$$
in which $ f$ is given in \Eq{fdef}.

We have also computed the two point function for $A_2$. At $\rtilde=0$
we again find
$$
\langle A_2^{i} A_2^{i}\rangle(p+\Pi) = -
\langle A_1^{i} A_1^{i}\rangle(p),
\EQN{A2propagain}
$$
where $\Pi=(\pi,\pi,\pi,\pi)$, as in section \use{sect.SCinEP}. However,
for $\rtilde\ne 0$, $A_2$ has a mass of the order of the cutoff in the
symmetric phase and also through the phase transition. Therefore
this particle does not appear in the physical spectrum.

Next we give results for the Wilson--Yukawa version of the
Eichten--Preskill model in the large $N$ limit, so that a comparison can
be made. We will set $\phi_2=0$ for simplicity, although we have
verified that the results for $A_2$ in the four fermion model are
reproduced.
We add a potential term to the lagrangian in \Eq{lagEPY} of the form
$$
V(\phi_1) = {g\over4}\lbrack \sum_a\phi_1^a\phi_1^a\rbrack^2,
\EQN{phipotential}
$$
thus introducing an additional coupling, $g$. In addition, we set $b=1$
and we rescale the couplings for this expansion:
$$
\ytilde = y\sqrt{N},\ \ \ \ \
\wtilde = w\sqrt{N},\ \ \ \ \
\gtilde = gN,        \EQN{Ytildedefs}
$$
keeping $\ytilde$, $\wtilde$, $\gtilde$ and $\kappa\equiv\kappa_1$ fixed.

The results for the Wilson--Yukawa model are similar to those given
above for the Eichten-Preskill model, so we will merely
state them, distinguishing quantities in this model from those
above by a $Y$ subscript.
The fermion propagator is a straightforward summation
of bubble diagrams for the self-energy of the fermion and is given by
$$
S_Y^{-1}(p) = \pmatrix{i\sum_\mu \sigmahat_\mu \sin p_\mu&
\sum_a\eta^a T^\dagger_a \Sigma_Y(p)\cr
\sum_a\eta^a T_a \Sigma_Y(p)&i\sum_\mu \sigmabar_\mu \sin p_\mu\cr},
 \EQN{Ypropparam}
$$
where
$$
\Sigma_Y(p) = \vtilde (\ytilde+\wtilde C(p)), \EQN{YSigmadef}
$$
in which $\vtilde$ comes from the vacuum expectation value of the scalar field:
$$
v = \langle\phi_1^1\rangle \equiv \vtilde\sqrt{N} \EQN{Yvdef}
$$
(again choosing the symmetry
breaking to occur in the $1$ direction).

Calculating the right hand side of \Eq{Yvdef} explicitly and setting it
equal to $v$ leads to an equation for
finding the critical surface between the symmetric and the broken phase.
This equation is similar to the expression in \Eq{criticalsurface},
except that it involves the four parameters $\ytilde$, $\wtilde$, $\kappa$
and $\gtilde$ rather than the two couplings of the previous model:
$$
1+\gtilde \vtilde^2 -8\kappa=
6\lbrack\ytilde^2 \IY0+2\ytilde\wtilde \IY1 + \wtilde^2 \IY2\rbrack,
\EQN{tadpoleequation}
$$
where
$$
\IY{n} = {16\over 3}\int_q {C^n(q)\over\sum_\mu \sin^2q_\mu +\Sigma_Y^2(p)},
\EQN{Intdef}
$$
similar to the previous case. The critical surface is found by allowing
$\vtilde\to0$ in this equation.

To calculate the boson propagators, let us define
$$
\phi_1^a = v\eta^1 + \xi^a,
\EQN{xsidef}
$$
where $\eta^1$ is the unit vector in the $1$ direction.
The propagator for the Higgs field $\xi^1$ is then
$$
\EQNalign{
\langle \xi^1\xi^1\rangle(p) =& {1\over
1 + 3\gtilde\vtilde^2-8\kappa + 2\kappa C(p) -{\cal H}^-_{Y}(p)},
\EQN{YHiggsprop}\cr
\noalign{\hbox{and the propagator for the Goldstone modes is}}
\langle \xi^j\xi^j\rangle(p) =& {1\over
1 + \gtilde\vtilde^2-8\kappa + 2\kappa C(p) -{\cal H}^+_{Y}(p)},
\;\; j=2,\ldots,10,
\EQN{YHiggsprop}\cr
}
$$
where
$$
{\cal H}^\pm_{Y}(p)= \ytilde^2 \HY0\pm(p) + 2\ytilde \wtilde \HY1\pm(p)
+ \wtilde^2\HY2\pm(p),
\EQN{HYdef}
$$
in which
$$
\HY{n}\pm(p) = 32\int_q\lbrack\half(C(q)+C(p-q))\rbrack^n
{
\sum_\mu\sin q_\mu\sin(q_\mu-p_\mu)\pm\Sigma_Y(q)\Sigma_Y(q-p)
\over
D_Y(q) D_Y(q-p)
},
\EQN{Hdef}
$$
and
$$
D_Y(q) = \sum_\mu \sin^2q_\mu + \Sigma_Y^2(q).
\EQN{DYdef}
$$
Finally the effective Yukawa coupling corresponding to
\Eq{effectiveYukawacoupling1} is
$$
y_{\scriptscriptstyle\rm Y}(p,q)
={\Sigma_Y(q)+\Sigma_Y(p+q)\over 2 v}.
\EQN{effectiveYukawacoupling1}
$$

Our last task is to map the two models into one another by
finding relations between the couplings. Indeed this can be done, and a
comparison of all quantities derived above results in the following
matching conditions:
$$
\EQNalign{
{\rtilde\over\wtilde}=
&{\vtilde\over f}={\rho\over\ytilde}, \EQN{matching1}\cr
\kappa=&\gtilde=0. \EQN{matching2}\cr
}
$$
With these conditions, all quantities derived above are identical for
the two models in the scaling region.
Of course, \Eq{matching2} reveals only that no
corresponding parameters to $\kappa$ and $g$ were included in the
lagrangian of the
Eichten--Preskill model. If such terms would be
included, matching conditions would be obtained for
them also. Thus we conclude that the two models give exactly the
same physics in the weak coupling
large $N$ limit, where we see explicitly the
presence of both a symmetric and a broken phase.

Aside from the details of the model and the field content, our calculation is
very similar to that performed in \Ref{Kutietal} in
relating the top quark condensate model and the standard model, with similar
results.

%%%#] Large N:
%%%#[ Conclusion:

\section{Conclusion}

When Eichten and Preskill originally presented their model, a clear
element of the scenario they envisaged for it to successfully produce a
continuum theory of (asymptotically free) chiral fermions entailed the
existence of a phase transition for which the fermion mass
was an order parameter, and over which no symmetry breaking occurred.
It was hoped that one could arrange through
generalized Wilson terms to give the doublers a mass, and leave the
lightest fermion massless. We have analyzed the model in several regions
of the phase diagram, and all indications are that no such phase
transition exists. Indeed, we do find phases with massive and massless
fermions, but always a broken phase appears in between.
In the symmetric phase with massive fermions (a paramagnetic
phase in strong Yukawa coupling, or PMS phase), bound states are formed
which pair up with the original chiral fields to form Dirac
representations, all of which are massive (although one massless Dirac
fermion can be arranged by tuning). The fermions remain massive across
the symmetry breaking phase boundary to the broken phase (ferromagnetic
or FM phase), and finally across the phase boundary to the second
symmetric phase (paramagnetic phase at weak Yukawa coupling, or PMW
phase), all fermions become massless, including the doublers. The
crucial ingredient for the failure of the emergence of a chiral theory
of fermions as originally imagined is the existence of the broken phase
separating the two symmetric phases. Through this broken phase the fermion
masses smoothly interpolate from being massive to vanishing, becoming
proportional to the order parameter of symmetry breaking as the second
phase boundary is approached.

Of course in order to conclude that the fermions are truly non-chiral,
one should also check that the right-handed partners to the left-handed
fields which are formed as bound states in the PMS phase couple with
equal strength to the gauge fields when gauge interactions are turned
on. The fact that both have the same charge with respect to the chiral
group would lead one to expect that this is the case, and indeed we have
checked this to be true explicitly in the presence of background gauge
fields, in a calculation similar to that presented for the modified
Smit--Swift model in \Ref{JanandUs}.

Through making contact with a Yukawa version of the original
Eichten--Preskill model, we see that this whole picture is rather
reminiscent of what happens in
the Smit--Swift model. Despite the essential difference
in the symmetry structure of the two models, the phase diagram appears
remarkably similar. This presumably indicates that the main aspect
responsible for the determination of the phase diagram is the method by
which the fermions obtain masses which is similar in both cases, rather
than the symmetry involved, or anything to do with anomalies. The
remainder of this section contains a more specific overview of the
calculations we have performed, and their implications for various
regions of the phase diagram.

Our calculations are primarily within the framework of $1/N$ expansions
for a model in which we have included $N$ copies of the original
fermions. We have studied both the original Eichten--Preskill model, and
a model with an explicit scalar field coupled to the fermions through
Yukawa couplings, and have argued through both universality and through
direct calculation in the large $N$ limit that the two produce the same
physics in the respective scaling regions. The first of these contains
two parameters, the four fermion coupling $\lambda$ and a four fermion
generalized Wilson coupling $r$. Through comparison with the second
model, two other relevant parameters are needed corresponding to a mass
and/or hopping term for a scalar excitation (which is a bound state in
the first model) and a self-interaction term for this field. In the
Yukawa version of the model, all these couplings are explicit, with a
Yukawa coupling $y$ corresponding to the four fermion coupling
$\lambda$, a Wilson--Yukawa coupling $w$ corresponding to $r$, and the
usual scalar hopping parameter $\kappa$ and four point coupling $g$.

In
the weak coupling region we have obtained an exact solution of both
models in the large $N$ limit, with matching conditions between the two
models. This calculation leads to the conclusion that in the weak
coupling region the fermions are massless, but when either $\kappa$ or
one of the Yukawa couplings $y$ or $w$ or a combination thereof (or
alternatively a combination of the four fermion couplings $\lambda$ or $r$)
is increased, a broken phase is reached in which the fermions all get a
mass proportional to the expectation value of the scalar field. (In terms
of the Eichten--Preskill model this is the Nambu--Jona-Lasinio phase
transition.)  These results do not appear to depend on the value of the
four point coupling for the scalar interaction although the large $N$
results are related to the weak coupling region.

In the strong coupling region, strong coupling techniques for the
Eichten-Preskill model tell us that the fermions are
massive\cite{EichPres} and through a study of the scalar excitation one
sees that as the coupling is decreased a broken phase is reached with a
phase transition occurring for a value of the coupling larger
than that for the Nambu--Jona-Lasinio phase transition. This provides clear
evidence that the two symmetric phases are separated by a broken phase
interpolating between the two. Large $N$ techniques can also be applied
to this region to compute the fermion determinant in the Yukawa model,
thereby  giving a partial solution at strong coupling in terms of an
effective action for the scalars. This effective action is strongly
peaked around a frozen scalar field, and mean field techniques apply.
Through this route one obtains an additional estimate for the value of
the critical coupling where the broken phase is reached, and it agrees
well with the estimate from strong coupling in the Eichten--Preskill
model.

We have also shown that in both the model as originally formulated by
Eichten and Preskill, and the Yukawa model we have introduced, a scalar
arises that is odd under CP transformations, and which has undesirable
properties. One way this problem could presumably be repaired is through
the addition of a term $-{\lambda\over 12}A^*A$ to the lagrangian in
\Eq{SClagEP} so that the scalar field $\phi_2$ is not needed in
\Eq{onepoint}. We have not investigated this possibility, as this change
will probably not alter our principal finding that a theory of chiral
fermions does not emerge.

Now where are possible loopholes in our conclusion that no theory of
asymptotically free chiral fermions emerges from the Eichten--Preskill
model? First our results are almost entirely within the framework of the
large $N$ expansion, and one could question whether something different
could happen for finite $N$. Although this remains possible, because of
the equivalence of the large $N$ expansion to first order perturbative
or strong coupling expansions, and because of the similarity of the
results qualitatively to those of other
models\cite{JanandUs}\cite{Usoneoverw}, we believe the results to be
qualitatively correct. Additionally, we have not studied the entire
phase diagram, but have limited ourselves to the positive $\kappa$ region.
In particular, we have presented no argument that prevents the two
phase boundaries from continuing below $\kappa=0$ to intersect at some
negative value of $\kappa$ thus becoming a critical surface between the
PMS and the PMW phases.  Although previous investigations in other models
\cite{Kutietal}\cite{Aachenphase} suggest that this will not occur,
it remains a possibility, however unlikely,
which in principle could be checked via numerical
simulations (in practice it could prove to be rather
difficult). Thus although we do not consider the chapter entirely
closed, we consider it highly unlikely that the Eichten--Preskill model
will produce a field theory of chiral fermions.

%%%#] Conclusion:
%%%#[ Acknowledgments:

\Acknowledgements

One of us (M.~G.) would like to thank E.~Eichten and J.~Kuti for some
useful discussions.
M.~G. and D.~P. have support from the Department of Energy and E.~R. is
supported by a Formaci\'on del Personal Investigador fellowship from the
Spanish government. Part of this research was carried out at the Aspen
Center for Physics, Aspen, Colorado, to which the authors are grateful.
\vfill

%%%#] Acknowledgments:
%%%#[ Appendix:

\appendix{A}{A Chiral representation for $SO(10)$\label{app.groups}}

The Clifford algebra in $11$ dimensions is generated by the $32$
dimensional hermitian gamma matrices satisfying
$$
\Gamma_a\Gamma_b + \Gamma_b\Gamma_a = 2 \delta_{ab}.
\EQN{Agammaalg}
$$
The hermitian generators of $SO(11)$ in this representation are given by
$$
M_{ab}= -{i\over 4}\lbrack\Gamma_a,\Gamma_b\rbrack.
\EQN{Agenerators}
$$
These satisfy the commutation relations
$$
\lbrack M_{ab},\Gamma_c\rbrack=
i\left(\delta_{bc}\Gamma_a-\delta_{ac}\Gamma_b\right).
$$
In limiting ourselves to $SO(10)$, generated by $M_{ab},a,b=1,
\ldots,10$, and the corresponding $10$ gamma matrices,
the matrix $\Gamma_{11}$ plays a role similar to $\gamma_5$ in four
dimensions. Thus we may define projectors
$$
P_{\pm} = \half\left(1\pm\Gamma_{11}\right),
\EQN{Aprojectors}
$$
satisfying
$$
\Gamma_a P_\pm = P_\mp\Gamma_a,\;\;\; a=1,\ldots,10.
\EQN{Agammaproj}
$$
Because the generators of $SO(10)$ are bilinears, \Eq{Agammaproj}
implies that they commute with $P_\pm$.
A convenient representation in terms of a direct product of Pauli matrices,
in which $\Gamma_{11}$ and hence $P_\pm$ are diagonal, is:
\def\psig#1#2{\sigma_{#2}}
\def\pone{\unity\>}
\def\sig#1#2{\times\psig{#1}{#2}}
\def\sone{\times\pone}
$$
\EQNalign{
\Gamma_1    =& \psig{1}{1}\sig{2}{1}\sig{3}{1}\sig{4}{1}\sig{5}{1},\cr
\Gamma_2    =& \psig{1}{2}\sig{2}{1}\sig{3}{1}\sig{4}{1}\sig{5}{1},\cr
\Gamma_3    =& \psig{1}{3}\sig{2}{1}\sig{3}{1}\sig{4}{1}\sig{5}{1},\cr
\Gamma_4    =& \pone    \sig{2}{2}\sig{3}{1}\sig{4}{1}\sig{5}{1},\cr
\Gamma_5    =& \pone    \sig{2}{3}\sig{3}{1}\sig{4}{1}\sig{5}{1},\cr
\Gamma_6    =& \pone    \sone     \sig{3}{2}\sig{4}{1}\sig{5}{1},\cr
\Gamma_7    =& \pone    \sone     \sig{3}{3}\sig{4}{1}\sig{5}{1},\cr
\Gamma_8    =& \pone    \sone     \sone     \sig{4}{2}\sig{5}{1},\cr
\Gamma_9    =& \pone    \sone     \sone     \sig{4}{3}\sig{5}{1},\cr
\Gamma_{10} =& \pone    \sone     \sone     \sone     \sig{5}{2},\cr
\Gamma_{11} =& \pone    \sone     \sone     \sone     \sig{5}{3}.\cr
}
$$
This representation is analogous to the chiral representation in four
dimensions, and $\Gamma_{11}$ can be written
$$
\Gamma_{11}=-i\Gamma_1\Gamma_2\cdots\Gamma_{10}.\EQN{Agammaeleven}
$$
The charge conjugation matrix $C$ is defined according to
$$
\EQNalign{
C\Gamma_a &= -\Gamma_a^T C,\cr
C^T=C^{-1}&=C^\dagger = -C,\EQN{Achargeconj}\cr
}
$$
and in the representation defined above, this matrix can be taken to be
$$
C=i\psig{1}{2}\sig{2}{3}\sig{3}{2}\sig{4}{3}\sig{5}{2}.
\EQN{Achargeconjexpl}
$$

%%%#] Appendix:

\vfill
\eject
\center{{\bf References}}
\bigskip
\ListReferences

%)]}
\bye